\documentclass[conference]{IEEEtran}
\usepackage{array}
\usepackage{multirow}

\usepackage[noadjust]{cite}
\usepackage{algorithm}
\usepackage{algorithmic}
\usepackage{amssymb}
\usepackage{bm}
\usepackage{amsmath}
\usepackage{cases} 
\usepackage{multirow}
\usepackage{booktabs}
\usepackage[caption=false,font=footnotesize]{subfig}
\usepackage{siunitx}
\usepackage{tabularx}
\usepackage{makecell}
\usepackage[table,xcdraw]{xcolor}

\ifCLASSINFOpdf
  \usepackage[pdftex]{graphicx}
  \graphicspath{{./}{./}}
  \DeclareGraphicsExtensions{.pdf,.jpeg,.png}
\else
  \usepackage[dvips]{graphicx}
  \graphicspath{{./}}
  \DeclareGraphicsExtensions{.eps}
\fi
\begin{document}

\title{Overlapped List Successive Cancellation Approach for Hardware Efficient Polar Code Decoder 
\vspace{-0.5em}}

\author{\IEEEauthorblockN{Tiben Che, Jingwei Xu and Gwan Choi}
\IEEEauthorblockA{Department of Electrical and Computer Engineering\\
Texas A\&M University, College Station, Texas 77840\\
Email: $\lbrace$ctb47321, xujw07, gchoi$\rbrace$@tamu.edu}}


%


\maketitle

\begin{abstract}
This paper presents an efficient hardware design approach for list successive cancellation (LSC) decoding of polar codes. 
By applying path-overlapping scheme, the $l$ instances of $(l>1)$ successive cancellation (SC) decoder for LSC with list size $l$ can be cut down to only one. 
This results in a dramatic reduction of the hardware complexity without any decoding performance loss. 
We also develop novel approaches to reduce the latency associated with the pipeline scheme. 
Simulation results show that with proposed design approach the hardware efficiency is increased significantly over the recently proposed LSC decoders. 
\end{abstract}

%
\IEEEpeerreviewmaketitle

 \section{Introduction}
\label{Introduction}
Recently, polar codes \cite{arikan2009channel} have received significant attention due to its capability to achieve the capacity of binary-input memoryless symmetric channels with low-complexity encoding and decoding schemes. 
E. Arikan in~\cite{arikan2009channel} presents a recursive cancellation way to successively accomplish decoding; and this method is referred to successive cancellation (SC). 
Also, N. Hussami et al. in~\cite{hussami2009performance} shows that the belief propagation (BP) can be applied as decoding algorithm.   
However, the decoding performances of both SC and BP are inferior to that of low density parity check (LDPC) codes. 
In order to make polar codes more competitive, the list SC (LSC) decoding algorithm is presented in~\cite{tal2011list}. 
By exploiting a larger range in the codeword tree, LSC significantly improves the decoding performance. 

Attracted by the potentials of LSC, a number of relevant hardware designs have been explored. 
In~\cite{balatsoukas2014hardware}, hardware LSC architectures of list sizes two and four are proposed with pointer memory technique, which can avoid the high complexity of likelihood copying.
In~\cite{lin2014efficient}, a hardware efficient architecture of LSC concatenated with cyclic redundancy check (CRC) is presented.
In~\cite{zhang2014hardware}, a hardware architecture of sub-optimal version of LSC decoding is introduced.
In~\cite{yuan2015low}, a LSC with multi-bit decision is discussed, which significantly reduces the decoding latency, and the corresponding hardware architecture is presented. 
All of aforementioned designs are using $l$ duplications of SC decoder for LSC decoder with list size $l$. 
Consequently, compared with SC decoder, the complexity of LSC increases from $n{\log} n$ to $l\cdot n{\log} n$, where $n$ and $l$ are the length of codeword and list size, respectively.
However, such complexity increasing makes all current existing LSC architectures are impractical for decoders with large list size.

This paper presents a hardware design approach for LSC decoding using path-overlapping to maximize hardware efficiency for optimal energy utility. 
Instead of using $l$ copies of SC decoder for LSC decoder, only one SC decoder used in our design. 
The calculations associated with each path occur simultaneously in the same decoder by carefully arranging the hardware configuration and scheduling of SC decoding. 
We arrange the LLR calculations of each path instantiated to occupy the decoder hardware stages serially in a streamlined fashion. 
This yields a significant reduction of hardware complexity. 
We also analyze and mitigate the latency overhead incurred in the path-overlapping scheme. 
Three approaches developed to reduce this overhead are: multi-decision LSC decoding, path-LLR-compute-ahead scheme and adaptive LSC decoding. 
The simulation results show that with proposed approach, the widely proposed LSC decoder can achieve a significantly higher hardware efficiency. 
For instance, for LSC decoder with code length $n=1024$ and list size $l=4$, at least 50\% hardware efficiency improvement achieved with proposed design approach, and the maximum improvement is up to around 130\%.

This paper is organized as follows. 
The relative background is reviewed in section~\ref{Background}. 
In following, the proposed approach is described in section~\ref{Proposed Design}. 
After that, the hardware efficiency performance and relevant analysis are presented in section~\ref{Comparison and Discussion}. 
Finally, this paper is concluded in section~\ref{Conclusion}.

\section{Background}
\label{Background}
\subsection{Polar Code}
\label{Polar Code}

As introduced in \cite{arikan2009channel}, a polar code is constructed by successively performing channel polarization.
Mathematically, polar codes are linear block codes of length $N = 2^n$. 
The transmitted codeword ${\bm{x}}\triangleq {(x_1,x_2,\cdots,x_N)}$ is computed by $\bm{x}=\bm{u}\bm{G}$ where $\bm{G=F^{\otimes m}}$, and $\bm{F^{\otimes m}}$ is the $m$-th Kronecker power of 
$\bm{F} = 
\begin{bmatrix}
1&0\\
1&1
\end{bmatrix}
$. 
Each row of $G$ is corresponding to an equivalent polarizing channel. 
For an $(N,k)$ polar code, $k$ bits that carry source information in $\bm{u}$ are called information bits. 
They are transmitted via the most $k$ reliable channels. 
While the rest $N-k$ bits, called frozen bits, are set to zeros and are placed at the least $N-k$ reliable channels. 
\begin{figure}[!tc]
\centering
\includegraphics[width=3in]{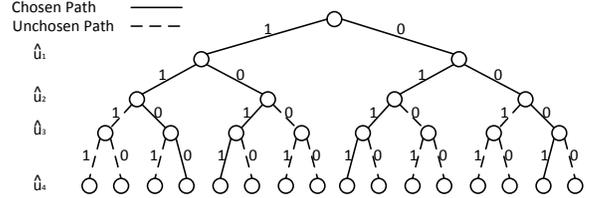}
\caption{An example of LSC decoding with list size 4 for $(8,4)$ polar code from codeword tree aspect}
\label{code_tree}
\end{figure} 

Polar codes can be decoded by recursively applying successive cancellation to estimate $\hat{u}_i$ using the channel output $y_{0}^{N-1}$ and the previously estimated bits $\hat{u}_{0}^{i-1}$. 
The calculation starts from channel output to the codeword, and is computed stage by stage. 
Polar code with length $n$ has ${\log} _{2}n$ stages. The previously estimated bits for intermediate stages are called partial sum.   
This decoding process of polar code can be regarded as the path searching in the code tree. 
SC decoding reserves only one survival path every layer.
If multiple paths are reserved in every layer, it is LSC decoding. The more paths survive, the higher chances the correct codeword can be found. 
~Fig.~\ref{code_tree} shows an example of LSC decoding with list size 4 for $(8,4)$ polar code from codeword tree aspect.

\subsection{Conventional architecture of LSC}
\label{Conventional architecture of LSC}

For the LSC algorithm, every information bit can derive two candidate paths, which are used to represent the decision of bit as $0$ or $1$. 
Each path has its own path metric which is corresponding to its survival probability. 
When performing the LSC decoding, $l$ paths are expanded to $2l$ paths for each estimated information bits.
Then the metrics of $2l$ paths are calculated to decide the $l$ survivals.
All the corresponding inner log likelihood ratios (LLRs) and partial sum of the reserved paths need to be kept along with $l$ paths as well.  
Finally, the $l$ paths are fed back to SC decoders and do all the steps again and again until the last information bit is decoded.   

Although all the LSC designs mentioned in Section~\ref{Introduction} have differences at some details, the main architecture are similar. 
Typically, for a LSC decoder, it has $l$ copies of SC decoders and one metrics computation units (MCU), one sorting module and three memory banks with respect to path metrics, current survival paths and LLRs and partial sums. 
The SC decoder consists of multiple processing units (PUs) with a tree architecture which consumes most of hardware resources. 
Such duplications of SC decoder yield a significant hardware redundancy of LSC decoder design. In our proposed design, we are trying to avoid such unnecessary redundancy.

\section{Proposed Approach}
\label{Proposed Design}
\begin{figure}[!tc]
\centering
\includegraphics[width=3in]{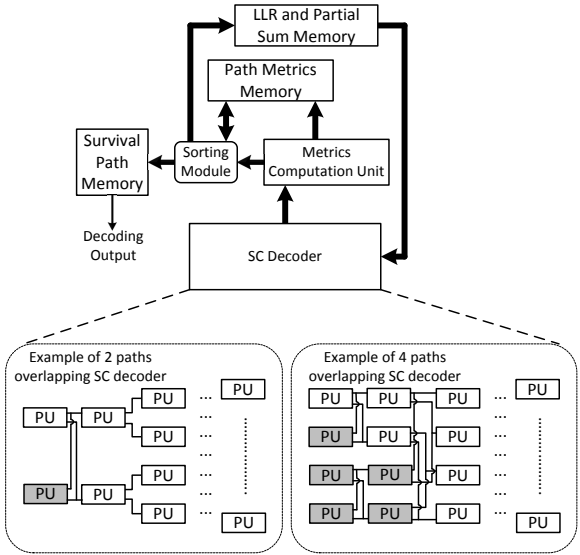}
\caption{The architecture of proposed design}
\label{proposed_design}
\end{figure} 
In this section we present our path-overlapping approach and discuss how performance optimization is carried out. 
Fig.~\ref{proposed_design} shows the architecture of proposed approach and the examples of the modified architecture of SC decoders associated with the list sizes two and four.  
Since the duplications of SC decoder involves the most hardware complexity, we removed all the copies and kept only one SC decoder. 
However, this modification of architecture does not mean that we just simply change parallel computing to a single-threaded lazy serial approach that computes one path at a time. 
Instead, every path is computed simultaneously in the decoding threads by judiciously utilizing the decoder hardware as follows: The processing timing of each path is overlapped with others in the pipeline arrangement. 
The architecture of SC decoder is modified to support this new paradigm. 
Since modifications are made only on architecture and scheduling plan, no decoding performance gain loss or change is incurred. 
The sorting module, MCU, and related memory components are compatible with other LSC decoders, and the partial sum generator is scheduled a similar way to be compatible with the path-overlapping SC decoder. 
Thus we do not discuss that in this paper. 
In the next subsections, the details of the scheme and the specific SC decoder are discussed.

 \subsection{Path-Overlapping Scheme and Relevant Analysis}
\label{Path-Overlapping Scheme and Relevant Analysis}


\begin{figure}[]
\centering
\subfloat[]{\includegraphics[width=2.7in]{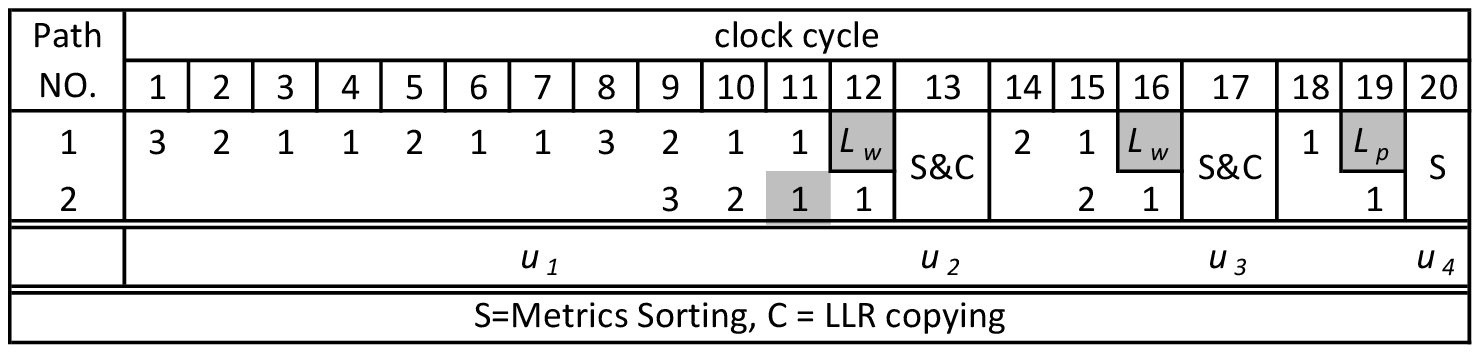}\label{path_2}}
\vfil
\subfloat[]{\includegraphics[width=2.7in]{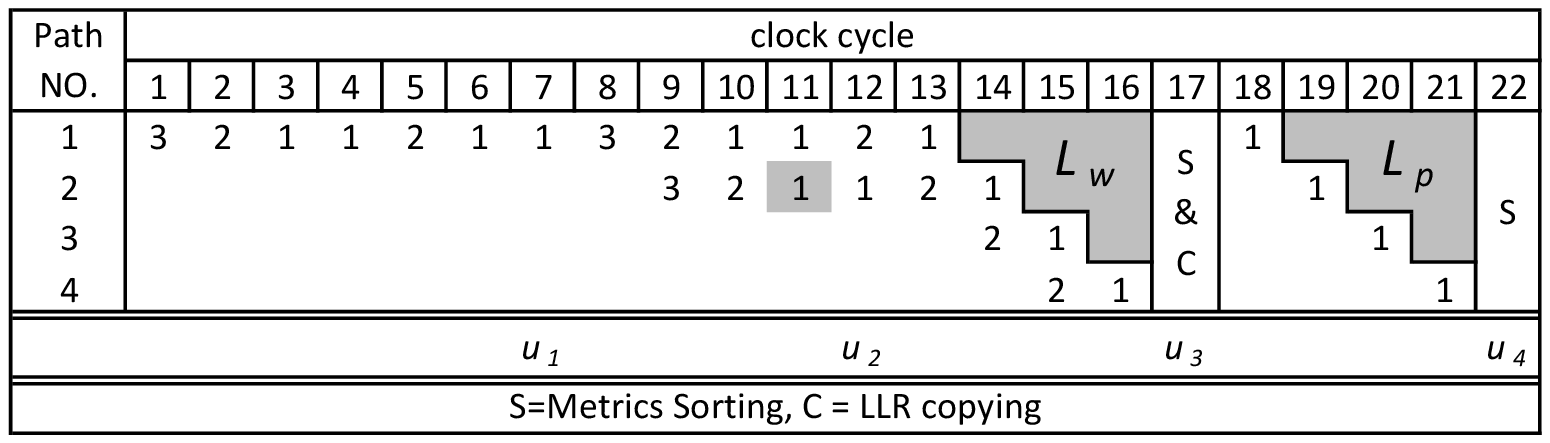}\label{path_4}}
\caption{Decoding schedule of the path-overlapping scheme for $(8,4)$ polar code with  \protect\subref{path_2} list size = 2  and \protect\subref{path_4} list size = 4} 
\label{Decoding schedule}
\end{figure}
Simultaneous processing approach is already presented in some SC decoders, and it is used for multiple frames in order to increase the throughput~\cite{zhang2013low}.
The SC decoder with tree architecture consists of multiple processing unites (PU) arranged like a binary tree. 
For every clock cycle, only one stage of PUs in the tree is activated. 
The basic idea of simultaneous processing approach is activating multiple decoding stages in one clock cycle by feeding in several frames in pipeline. This means that each frame comes into the decoder with one clock cycle delay.  

%
%
Stemming from above idea, we realize that the duplications of SC decoder in conventional LSC decoder is unnecessary. All the paths can be fed into the same decoder in pipelined fashion. Different stages in the single SC decoder can process different paths simultaneously. 
Computations of successive paths are overlapped in temporal with only one clock cycle delay. 
However, the decoding scheme is not exactly the same as multiple frames overlapping SC decoder.   
Fig.~\ref{path_2} and Fig.~\ref{path_4} show the decoding schedule of two and four path-overlapping scheme, respectively.
The number means current activated stages, and the duplicated stage is marked with gray. 
According to~\cite{zhang2013low}, if a SC decoder is with $l$ path-overlapping scheme, where $l\leq(2^i-1)$, it can be constructed by duplicating $(2^{i-1}-1)$ stages, where the index starts from the information bits side with respect to the tree architecture. 
The duplication plan is also presented in Fig.~\ref{proposed_design}.  
Noticeably in Fig.~\ref{path_4} there is only one duplication of stage one, which is not the same as what presented in Fig.~\ref{proposed_design}. 
This is because the number of copies in Fig~\ref{proposed_design} are the minimum requirement for all the case. The actual requirement is decided by the code length and rate. Fig.~\ref{path_4} is just a certain case only one stage duplication is needed for four path-overlapping scheme.

Such architecture significantly reduces hardware complexity. 
Another advantage of proposed approach is that it can reduce the critical path length of decoder. 
Usually, the critical path lies in the sorting block. 
For conventional LSC decoder, the sorting block is composed of staged combination logic.
Even for very small list size, e. g. list = 4, the critical path is much longer than any other module. 
With proposed approach, since each path metrics comes with pipeline arrangement, naturally, the sorting block is designed as a pipeline module which has a shorter critical path than that of combination logic for the same list size. This means, by applying proposed approach, LSC decoder can run at a much higher frequency.       

Although proposed approach can achieve a higher frequency compared with the conventional LSC decoder, there are some additional clock cycles introduced. 
These consist of two parts. 
The first part is the path pipeline latency $L_p$. Since all the paths are fed into decoder with one clock cycle delay, for the LSC with list size $l$, $L_p~=~(l-1)$. 
The second part is path waiting latency $L_w$. After the number of path extending to the maximum, the pipeline processing has to suspend when estimating the newly generated information bit since the decoder needs to wait for all the paths to finish before commencing metric sorting and LLR copying. 
This waiting period is referred to as pipeline stalling.
The waiting time is equal to $L_p$.
Thus, for the list size $l$ LSC with respect to $(n,k)$ polar code, $L_w~=~(k-{\log} _{2}l-1)\cdot (l-1)$.
Thus, the total latency overhead introduced by path-overlapping scheme $L_m$ can be calculated by: 
\begin{equation}\label{TotalD}
L_m~=~L_w+L_p~=~(k-{\log} _{2}l)\cdot (l-1).
\end{equation} 

This design approach can be applied to any current existing LSC decoders. 
It significantly reduce the hardware complexity by eliminating redundant instances, and it incurs few additional clock cycles to achieve the improvement.
Thus, it is difficult to evaluate such design approach merely in term of the usage of hardware resource or the latency. 
Thus we introduce the hardware efficiency (HE) metric which is noted as $e$ to measure the performance of proposed approach. The $e$ is defined as: $e~=~Throughput/Area$.   

From~Eq.~(\ref{TotalD}), we can tell that the latency overhead would significantly aggregate with either list size or code rate, which can significantly diminish the $e$. In order to achieve a high $e$ with proposed approach, the latency overhead must be reduced to an acceptable level. 
In the next sections, we will present three approaches aimed at decreasing the latency overhead.  



\subsection{Latency Reduction via Multi-Decision List SC Decoding}
\label{Latency Reducetion with Mutil-Desicion List SC Decoding}
The first part of Eq.~(\ref{TotalD}) corresponds to the path waiting latency. 
For every instance of estimating an information bit, the pipeline processing has to suspend until all the paths finish calculations. This provides an observation that if the times of estimating the information bit can be reduced, the $L_w$ will decrease significantly.

Multi-decision is an approach of estimating $m$ bits $(m>1)$ instead of just one at the same time. 
It helps to reduce the number of estimations. Many approaches can be regarded as multi-decision~\cite{yuan2015low}~\cite{xiong2014symbol}~\cite{che2015tc}~\cite{sarkis2015unrolled}.
Generally, they can be classified into two types. 
The first type is referred to as regular mutil-decision decoder; it estimates $m$ bits $(m>0)$ every time. 
Most of current multi-decision decoders belong to this type~\cite{yuan2015low}~\cite{xiong2014symbol}.    
The second type is called irregular mutil-decision decoders; the number of bits estimated every time is not fixed. 
Currently, only the list fast-SSC decoder~\cite{sarkis2015unrolled} belong to this type. 
It simplifies the SC decoding by finding certain pattern in the codewords. 
Such subcodes with certain pattern also refer to constituent codes.
The number of bits estimated every time is corresponding to the size of constituent code. 
Besides, the distribution of constituent codes irregularly change along with code rate. 

\begin{figure}[!tc]
\centering
\includegraphics[width=3in]{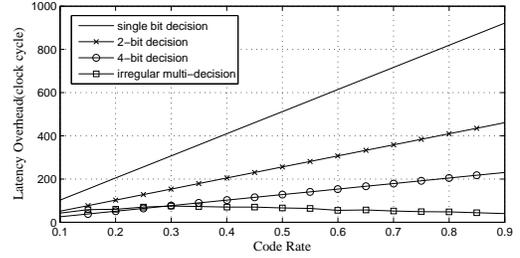}
\caption{latency overhead for different scheme}
\label{delay_multi}
\end{figure}

For path-overlapping LSC decoder with mutil-decision, $L_m$ can be further reduced to $L_m~=~\alpha\cdot (l-1)$. 
For $m$ bits regular mutil-decision, $\alpha~=~\lceil (k-{\log} _{2}l)/m \rceil. $ 
For irregular mutil-decision, $\alpha~=~S-{\log} _{2}l$ where $S$ is the total number of constituent codes which irregularly changes along with code rate. 
Fig.~\ref{delay_multi} shows the latency overhead of different schemes for LSC decoder with code length $n=1024$ and list size $l=4$. 
We can see that all the mutil-decision schemes can significantly reduce latency overhead, and as increasing of code rate, the irregular mutil-decision scheme can still keep a very low latency overhead.

\subsection{Latency Reduction via Path-LLR-Compute-Ahead Scheme}
\label{Latency Reduction with Path-LLR-Compute-Ahead Scheme}

\begin{figure}[!tc]
\centering
\includegraphics[width=2.7in]{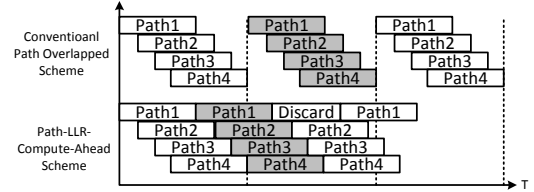}
\caption{decoding schedule of path-LLR-compute-ahead scheme}
\label{path_ahead}
\end{figure}

Besides reducing the number of estimations, the other approach to decrease latency overhead is by avoiding the pipeline stalling. 
This can be done via path-LLR-compute-ahead scheme (PLCAS). 
Fig.~\ref{path_ahead} shows this decoding schedule. 
A single bar means the decoding process between estimations of two successive information bits. 
When pipeline stalling happens in one path, instead of waiting, current path can do a pre-estimated between two candidates ($0$ and $1$) which it solely generates without suspension. 
The pipeline processing continues with the one with larger metrics and keeps the other to compared with the next coming paths. 
If more suitable paths are found later, the previous computed ones are discarded. 
With this scheme, the $L_m$ for the best case is equal to pipeline latency $L_p$, which means the entire processing is handled without any stalling, and the $L_m$ for the worst case is equal to simple path-overlapping scheme. 
 
\subsection{Latency Reduction via Adaptive LSC Decoding}
\label{Latency Reduction with Partial Parallel Architecture}

In Eq.~(\ref{TotalD}), the second part of the formulation is equal to the $L_p$. 
It is determined by the number of paths set in the pipeline. 
This makes the latency overhead increas linearly with respect to the list size $l$. 
If we can decrease the value, the latency overhead can be significantly reduced. 
Typically, $L_p$ is fixed for a LSC with given length. However, by applying adaptive LSC algorithm~\cite{7094411}, the $L_p$ is allowed to change on the fly according to current metrics of each path.
The list size would decrease along the decoding processing, which also means the latency overhead would get reduction.  

In~\cite{7094411}, basic hardware architecture is also proposed. 
Even though the list size would decrease along the decoding processing, the architecture proposed in~\cite{7094411} still needs $l$ copies of SC decoder for its initial status. 
The usage of hardware resource is same as regular LSC decoder. 
Proposed approach can exploit the metric of adaptive LSC decoder via cutting down the unnecessary hardware complexity. 
With proposed approach there is no redundant hardware even when the list size decrease. 
Such property allows adaptive LSC decoder to benefit more in term of $e$. 
This will be shown in section.~\ref{Comparison and Discussion}.        

\section{Performance and Analysis}
\label{Comparison and Discussion}

\begin{figure}[!htc]
\centering
\includegraphics[width=3.5in]{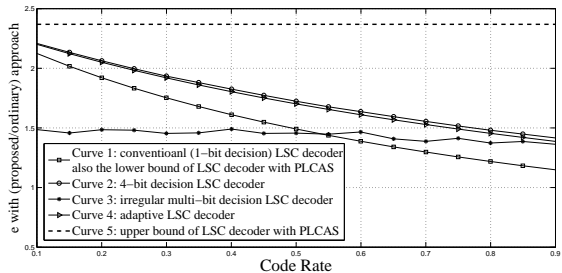}
\caption{the improvement of $e$ with proposed design approach}
\label{hardware_efficiency}
\end{figure}

Fig.~\ref{hardware_efficiency} shows the improvement of $e$ with proposed design approach for widely proposed LSC decoders with code length $n=1024$ and list size $l=4$. 
The x-axis is the rate of polar code, and the y-axis is the ratio of $e$ with proposed approach over $e$ with ordinary approach. The $e$ with ordinary approach for a given LSC decoder has a consistent value. 
We apply proposed approach to four types of LSC decoder.
They are conventional LSC decoder which also is regarded as 1-bit decision LSC decoder, 4-bit decision LSC decoder, irregular multi-bit decision decoder and the adaptive LSC decoder. 
We also calculated the upper and lower bound of the $e$ improvement with PLCAS.     
These simulations are based on the decoders described in~\cite{balatsoukas2014hardware},~\cite{yuan2015low},~\cite{sarkis2015unrolled} and~\cite{che2015tc}, the related synthesis results and the analysis we made in the previous sections.

In Fig.~\ref{hardware_efficiency}, all the curves are beyond the ratio of one, which means with the proposed approach, all the decoders are able to achieve a better hardware efficiency. 
According to curve 1 and curve 2, the hardware efficiency of regular decision decoder, 1-bit and 4-bit decision decoder, is decreasing alone with the code rate increasing. 
This is because the latency overhead is larger at higher code rate.
Besides, the regular multi-bit (4-bit) decoder achieves more improvement of $e$ than that of conventional (1-bit) decoder, which is due to the latency reduction as we described in section~\ref{Latency Reducetion with Mutil-Desicion List SC Decoding}. This can easily derive that for $n-bit$-decision regular decoder, the bigger the $n$, the more the improvement of $e$ can be achieved with proposed approach. 
Curve 1 and 5 indicate the range of the $e$ improvement with PLCAS. The actual value depends on the channel outputs and channel quality.  
According to curve 4 and curve 1, we can tell that the adaptive LSC help proposed approach to dramatically increase the hardware efficiency. Such increasing benefits from the decreasing of latency overhead as we analyze in section~\ref{Latency Reduction with Partial Parallel Architecture}. 
Another very interesting phenomenon is about the improvement of irregular multi-bit decision (list fast-SSC decoder). The gain of $e$ does not change too much with code rate varying. This is because the latency overhead of irregular multi-bit decision decoder does not linearly change along with coder rate. 
The average improvement of irregular multi-bit decision is less than that of regular one. 
This is due to the inherent latency of irregular LSC decoder is already very low~\cite{che2015tc}. 

Noticeably all the improvements are calculated based on the assumption that the maximum frequency of decoder with proposed approach or ordinary approach are the same. However, according to the analysis in section~\ref{Path-Overlapping Scheme and Relevant Analysis}, the maximum frequency of decoder with proposed approach should be higher, which indicates that the improvements of $e$ in Fig.~\ref{hardware_efficiency} should be even more in practice. 
Additionally, all the approaches mentioned above are not conflicting with each other. Using multiple approaches together can further increase the hardware efficiency. 
The above mentioned properties indicate that proposed approach can measurably contain the hardware complexity associated with large scale LSC decoder implementation.
\section{Conclusion}
\label{Conclusion}

This paper presents a novel design approach to improve the hardware efficiency of LSC decoder via path-overlapping scheme. The details of design approach and three strategies to reduce the latency overhead are also presented. The numerical results show that the conventionally used LSC decoders can significantly achieve a higher hardware efficiency using the proposed approach.   

\bibliographystyle{IEEEtran}
\bibliography{IEEEabrv}

\begin{thebibliography}{10}
\providecommand{\url}[1]{#1}
\csname url@samestyle\endcsname
\providecommand{\newblock}{\relax}
\providecommand{\bibinfo}[2]{#2}
\providecommand{\BIBentrySTDinterwordspacing}{\spaceskip=0pt\relax}
\providecommand{\BIBentryALTinterwordstretchfactor}{4}
\providecommand{\BIBentryALTinterwordspacing}{\spaceskip=\fontdimen2\font plus
\BIBentryALTinterwordstretchfactor\fontdimen3\font minus
  \fontdimen4\font\relax}
\providecommand{\BIBforeignlanguage}[2]{{%
\expandafter\ifx\csname l@#1\endcsname\relax
\typeout{** WARNING: IEEEtran.bst: No hyphenation pattern has been}%
\typeout{** loaded for the language `#1'. Using the pattern for}%
\typeout{** the default language instead.}%
\else
\language=\csname l@#1\endcsname
\fi
#2}}
\providecommand{\BIBdecl}{\relax}
\BIBdecl

\bibitem{arikan2009channel}
E.~Arikan, ``Channel polarization: A method for constructing capacity-achieving
  codes for symmetric binary-input memoryless channels,'' \emph{Information
  Theory, IEEE Transactions on}, vol.~55, no.~7, pp. 3051--3073, 2009.

\bibitem{hussami2009performance}
N.~Hussami, S.~B. Korada, and R.~Urbanke, ``Performance of polar codes for
  channel and source coding,'' in \emph{Information Theory, 2009. ISIT 2009.
  IEEE International Symposium on}.\hskip 1em plus 0.5em minus 0.4em\relax
  IEEE, 2009, pp. 1488--1492.

\bibitem{tal2011list}
I.~Tal and A.~Vardy, ``List decoding of polar codes,'' in \emph{Information
  Theory Proceedings (ISIT), 2011 IEEE International Symposium on}.\hskip 1em
  plus 0.5em minus 0.4em\relax IEEE, 2011, pp. 1--5.

\bibitem{balatsoukas2014hardware}
A.~Balatsoukas-Stimming, A.~J. Raymond, W.~J. Gross, and A.~Burg, ``Hardware
  architecture for list successive cancellation decoding of polar codes,''
  \emph{Circuits and Systems II: Express Briefs, IEEE Transactions on},
  vol.~61, no.~8, pp. 609--613, 2014.

\bibitem{lin2014efficient}
J.~Lin and Z.~Yan, ``Efficient list decoder architecture for polar codes,'' in
  \emph{Circuits and Systems (ISCAS), 2014 IEEE International Symposium
  on}.\hskip 1em plus 0.5em minus 0.4em\relax IEEE, 2014, pp. 1022--1025.

\bibitem{zhang2014hardware}
C.~Zhang, X.~You, and J.~Sha, ``Hardware architecture for list successive
  cancellation polar decoder,'' in \emph{Circuits and Systems (ISCAS), 2014
  IEEE International Symposium on}.\hskip 1em plus 0.5em minus 0.4em\relax
  IEEE, 2014, pp. 209--212.

\bibitem{yuan2015low}
B.~Yuan and K.~Parhi, ``Low-latency successive-cancellation list decoders for
  polar codes with multibit decision,'' 2015.

\bibitem{zhang2013low}
C.~Zhang and K.~Parhi, ``Low-latency sequential and overlapped architectures
  for successive cancellation polar decoder,'' \emph{Signal Processing, IEEE
  Transactions on}, vol.~61, no.~10, pp. 2429--2441, 2013.

\bibitem{xiong2014symbol}
C.~Xiong, J.~Lin, and Z.~Yan, ``Symbol-based successive cancellation list
  decoder for polar codes,'' in \emph{Signal Processing Systems (SiPS), 2014
  IEEE Workshop on}.\hskip 1em plus 0.5em minus 0.4em\relax IEEE, 2014, pp.
  1--6.

\bibitem{che2015tc}
T.~Che, J.~Xu, and G.~Choi, ``Tc: Throughput centric successive cancellation
  decoder hardware implementation for polar codes,'' \emph{arXiv preprint
  arXiv:1504.06247}, 2015.

\bibitem{sarkis2015unrolled}
G.~Sarkis, P.~Giard, A.~Vardy, C.~Thibeault, and W.~J. Gross, ``Unrolled polar
  decoders, part ii: Fast list decoders,'' \emph{arXiv preprint
  arXiv:1505.01466}, 2015.

\bibitem{7094411}
C.~Zhang, Z.~Wang, X.~You, and B.~Yuan, ``Efficient adaptive list successive
  cancellation decoder for polar codes,'' in \emph{Signals, Systems and
  Computers, 2014 48th Asilomar Conference on}, Nov 2014, pp. 126--130.

\end{thebibliography}
%

\end{document}